\RequirePackage{fix-cm}

\documentclass[smallextended]{svjour3}

\smartqed
\usepackage{dsfont, amsfonts, amssymb, graphicx}
\usepackage{bm}
\usepackage{hyperref}
\hypersetup{colorlinks=false}
\DeclareMathAlphabet{\mathpzc}{OT1}{pzc}{m}{it}

\journalname{General Relativity and Gravitation}

\begin{document}

\title{Acoustic black holes: massless scalar field analytic solutions and analogue Hawking radiation}

\titlerunning{Acoustic black holes}

\author{H. S. Vieira \and V. B. Bezerra}

\institute{H. S. Vieira \at Departamento de F\'{i}sica, Universidade Federal da Para\'{i}ba, Caixa Postal 5008, CEP 58051-970, Jo\~{a}o Pessoa, PB, Brazil\\Centro de Ci\^{e}ncias, Tecnologia e Sa\'{u}de, Universidade Estadual da Para\'{i}ba, CEP 58233-000, Araruna, PB, Brazil\\\email{horacio.santana.vieira@hotmail.com} \and V. B. Bezerra \at Departamento de F\'{i}sica, Universidade Federal da Para\'{i}ba, Caixa Postal 5008, CEP 58051-970, Jo\~{a}o Pessoa, PB, Brazil\\\email{valdir@fisica.ufpb.br}}

\date{Received: date / Accepted: date}

\maketitle

\begin{abstract}
We obtain the analytic solutions of the radial part of the massless Klein-Gordon equation in the spacetime of both three dimensional rotating and four dimensional canonical acoustic black holes, which are given in terms of the confluent Heun functions. From these solutions, we obtain the scalar waves near the acoustic horizon. We discuss the analogue Hawking radiation of massless scalar particles and the features of the spectrum associated with the radiation emitted by these acoustic black holes.
\keywords{Massless Klein-Gordon equation \and Rotating acoustic black hole \and Canonical acoustic black hole \and Confluent Heun function \and Black body radiation}
\PACS{02.30.Gp \and 04.20.Jb \and 04.70.-s \and 04.80.Cc \and 47.35.Rs \and 47.90.+a}
\end{abstract}
%
%
\section{Introduction}
In a seminal work, Unruh \cite{PhysRevLett.46.1351} showed that under certain conditions, the equation of motion describing the propagation of sound modes (phonons) on a background hydrodynamic flow, which undergoes a subsonic-supersonic transition, can be written in the same form of the Klein-Gordon equation for a massless scalar field minimally coupled to an effective Lorentzian geometry containing a sonic horizon. This means that this physical system may be considered analogous to astrophysical black holes and as a consequence should be used, in principle, to understand the physics of black holes as well as to realize experiments in laboratory to test some of their properties.

The existence of an event horizon in the hydrodynamic analogue suggests that an interesting phenomenon can be produced which consists in the emission of a thermal flux of phonons, whose temperature is proportional to the gradient of the velocity field at the acoustic horizon, termed analogue Hawking radiation, similar to the radiation emitted by astrophysical black holes, the well know Hawking radiation \cite{CommunMathPhys.43.199,PhysRevD.82.044013,PhysLettB.692.61,PhysLettB.697.398,EurophysLett.109.60006}. Thus, using this analogy, it is possible, in principle, to experimentally verify the analogue Hawking radiation emitted by acoustic black holes, and assuming that the physics which leads to Hawking radiation should be the same of its analogue, we can get some informations about a phenomenon which originates from the combination of quantum mechanics and general relativity, but now at a purely classical level.

It is difficult to observe the Hawking radiation emitted by astrophysical black holes due to the fact that Hawking temperature is seven orders of magnitude smaller than the temperature of the Cosmic Microwave Background (CMB) radiation, for a Schwarzschild black hole with a mass equivalent to the solar mass. Thus, the experimental verification of Hawking radiation emitted by astrophysical black holes could be possible only if these objects have mass much smaller than the solar mass. On the other hand, for acoustic black holes in condensed matter, the analogue Hawking temperature can be of about $10^{-6}$K to $10^{-7}$K (for a review, see \cite{LivingRevRelativity.8.12} and references therein), which is too low to be measured in laboratory, but certainly these values will be measured in a near future. The fact that Hawking radiation from astrophysical black holes has an analogue in hydrodynamic systems has stimulated the realization of several experiments \cite{NewJPhys.10.053015,PhysRevLett.106.021302} and suggested new experiments involving water waves \cite{PhysRevD.90.044033}.

Hawking radiation is an effect of kinematical origin which appears in the scope of general relativity connected with the existence of event horizons, and therefore, has no relation with aspects of the dynamic of the Einstein equations. Thus, if we have an analogue system which exhibits an event horizon, we expect the existence of an analogue of Hawking radiation as stated before. This means that this radiation is a more general phenomenon which occurs whenever an event horizon exists, and therefore, in different analogous gravitational systems. The fact that the temperatures involved in analogue models are very low suggests the realization of experiments involving ultra cold systems, such as Bose-Einstein condensates \cite{PhysRevLett.85.4643,PhysRevLett.91.240407,NewJPhys.13.063048,PhysRevLett.105.240401}, superfluid helium \cite{PhysRevD.58.064021}, superconductores \cite{JHighEnergyPhys.06.087,IntJModPhysD.21.1250038}, polariton superfluid \cite{PhysRevB.86.144505} and degenerate Fermi gas \cite{PhysRevLett.94.061302}. Otherwise, due to the general character of the Hawking radiation and the real possibility to detect its analogue, many other systems have been investigated \cite{PhysRevLett.107.149401,PhysRevLett.107.149402,PhysRevD.85.084014,EurPhysJPlus.127.78,PhysRevA.78.063804,PhysRevA.80.065802,EurophysLett.92.14002,PhysRevB.84.233405,NewJPhys.13.085005,NaturePhys.10.864}.

The astrophysical black holes and the corresponding analogue gravity models have many features in common, even possessing different dynamics. As for example, the dynamics of the acoustic black holes are governed by equations of fluid mechanics, while the dynamics of the astrophysical black holes are obtained from Einstein's equations. The fact that these different kinds of black holes possess some fundamental properties in common, associated with the possibility to detect the acoustic black holes in laboratory, which can help us to better understand the physics of a gravitational black hole, motivated a lot of investigation in which concerns the physics of acoustic black holes \cite{PhysRevD.83.124016,PhysRevD.83.124047,NewJPhys.10.103001,PhysRevD.92.024043,ClassQuantumGrav.18.1137,GenRelGrav.34.1719,JHighEnergyPhys.04.030,JPhysBAtMolOptPhys.45.163001,PhysRevA.78.021601,PhysRevD.90.104015,PhysRevD.81.124013,PhysRevD.79.064014,PhysRevD.70.124006,PhysRevD.51.2827,PhysRevD.44.1731,PhysRevD.48.728}.

A particular solution \cite{ClassQuantumGrav.20.3907} of the massless Klein-Gordon equation in the rotating acoustic background was obtained using a tortoise coordinate, in two distinct regions, namely, near the horizon and very far from it, with the proposal to examine the phenomenon of superresonance. In order to investigate the superresonant scattering of acoustic disturbances from a rotating acoustic black hole it was found an approximate solution \cite{ClassQuantumGrav.20.2929}, in the low-frequency limit, which is given in terms of hypergeometric functions. This solution, with appropriate boundary conditions, was also used to compute the quasinormal modes \cite{PhysLettB.617.174} associated with an acoustic rotating black hole.

In our paper we obtain the analytic solutions of the Klein-Gordon equation for a massless scalar field for both rotating and canonical acoustic black holes, valid in the whole space, which means between the acoustic event horizon and infinity. Furthermore, these solutions are valid for all frequencies. In this sense, we extend the range in which the solutions are valid as compared with the ones obtained by Basak and Majumdar \cite{ClassQuantumGrav.20.2929}, for the rotating acoustic black hole and additionaly the solution of the Klein-Gordon equation in the canonical acoustic black hole background was obtained. They are given in terms of solutions of the Heun equations \cite{Slavyanov:2000}. Using the radial solution which is given in terms of the confluent Heun functions and taking into account their properties, we study the analogue Hawking radiation of massless scalar particles.

In order to study Hawking radiation, diferrent approaches have been used. As examples, we can mention the following: (i) Hartle-Hawking \cite{PhysRevD.13.2188} which used analytic continuation; (ii) Christensen-Fulling which is based on the trace anomaly or conformal anomaly \cite{PhysRevD.15.2088}; (iii) Robinson-Wilczek which takes into account the anomaly cancellation \cite{PhysRevLett.95.011303}; (iv) Tunneling method \cite{PhysRevLett.85.5042} which uses the semiclassical WKB method, among others. Using these different approaches, Kim and Shin \cite{JHighEnergyPhys.07.070} considered the analogue Hawking radiation from the rotating acoustic black hole using the viewpoint of anomaly cancellation method. Ram\'{o}n \textit{et al.} \cite{IntJModPhysA.25.1463} presented the analogue Hawking radiation derivation from the canonical acoustic black hole using the anomaly and tunneling mechanism approaches. Zhang \textit{et al.} \cite{PhysLettB.698.438} studied Hawking radiation from a rotating acoustic black hole using the analytical continuation method \cite{PhysLettB.694.149}.

In this paper, we use just the radial part of the analytic solution of the Klein-Gordon equation, from which we construct the solutions near the acoustic horizons appropriate to examine Hawking radiation.

This paper is organized as follows. In the section 2, we introduce the metric that corresponds to a rotating acoustic black hole, and some relevant elements to study the Hawking radiation. We also present the analytic solution of the Klein-Gordon equation in this background, taking into account only the radial part. This solution is used to construct the ingoing and outgoing waves, near the acoustic horizon. We extend the waves solutions from outside to inside of the rotating acoustic black hole, and we derive a black body radiation spectrum. In the section 3, we do similar calculations in the background of the canonical acoustic black hole, and we derive the corresponding black body radiation spectrum. Finally, in section 4, we present our conclusions.
%
%
\section{Rotating acoustic black hole}
The solution which describes a (2+1)-dimensional rotating acoustic black hole was obtained by Visser \cite{ClassQuantumGrav.15.1767} and corresponds to an analogue model associated with a smoothly rotating draining fluid flow with a sink at the origin. Note that this fluid flow is constant in time and cylindrically symmetric, and therefore, represents a vortex line aligned along z-axis, without vorticity and being barotropic and inviscid. The acoustic metric appropriate to represent the effective Lorentzian geometry of this draining bathtube idealized model is given by \cite{ClassQuantumGrav.15.1767}
\begin{equation}
ds^{2}=-c^{2}\ dt^{2}+\left(dr-\frac{A}{r}dt\right)^{2}+\left(r\ d\phi-\frac{B}{r}dt\right)^{2}\ ,
\label{eq:metrica_draining_bathtub}
\end{equation}
where $c$ is the speed of sound which is constant throughout the fluid flow. The properties of this metric become evident by changing the coordinates, defined by
\begin{equation}
dt \rightarrow dt+\frac{|A|r}{c^{2}r^{2}-A^{2}}dr\ ,
\label{eq:relacao1}
\end{equation}
\begin{equation}
d\phi \rightarrow d\phi+\frac{B|A|r}{r\left(c^{2}r^{2}-A^{2}\right)}dr\ .
\label{eq:relacao2}
\end{equation}
Using these new coordinates and rescaling time coordinate by $c$, the line element that describes a rotating acoustic black hole can be written in the following form
\begin{equation}
ds^{2}=-\frac{1}{r^{2}}\left(\Delta-\frac{B^{2}}{c^{2}}\right)dt^{2}+\frac{r^{2}}{\Delta}dr^{2}+r^{2}\ d\phi^{2}-2\frac{B}{c}d\phi\ dt\ ,
\label{eq:metrica_draining_bathtub_Kerr}
\end{equation}
where
\begin{equation}
\Delta=r^{2}-\frac{A^{2}}{c^{2}}\ .
\label{eq:Delta_draining_bathtub_Kerr}
\end{equation}

Taking $g_{00}$ in Eq.~(\ref{eq:metrica_draining_bathtub_Kerr}) equal to zero, we obtain the radius of the ergosphere, which is given by
\begin{equation}
r_{e}=\frac{\sqrt{A^{2}+B^{2}}}{c}\ .
\label{eq:ergosphere}
\end{equation}
It is worth calling attention to the fact that the sign of $A$ is irrelevant in defining the ergosphere and ergo-region, since does not matter if the vortex core is a source or a sink. The metric has a (coordinate) singularity, that is, the acoustic event horizon forms once the radial component of the fluid velocity exceeds the speed of sound, which occurs at
\begin{equation}
r_{h}=\frac{|A|}{c}\ .
\label{eq:horizon}
\end{equation}
Here, the sign of $A$ must be taken into account. For $A < 0$ we are dealing with a future acoustic horizon, that is, an acoustic black hole; while for $A > 0$ we are dealing with a past event horizon, that is, an acoustic white hole.

Hence, from Eq.~(\ref{eq:Delta_draining_bathtub_Kerr}), we have that the acoustic horizon surface of the rotating acoustic black hole is obtained from the condition
\begin{equation}
\Delta=(r-r_{+})(r-r_{-})=0\ ,
\label{eq:superficie_hor_draining_bathtub_Kerr}
\end{equation}
whose solutions are
\begin{equation}
r_{+}=r_{h}\ ,
\label{eq:sol_padrao_draining_bathtub_Kerr_1}
\end{equation}
\begin{equation}
r_{-}=-r_{h}\ ,
\label{eq:sol_padrao_draining_bathtub_Kerr_2}
\end{equation}
where $r_{h}$ is given by Eq.~(\ref{eq:horizon}), and $r_{\pm}$ correspond to the acoustic events or Cauchy horizons of the rotating acoustic black hole.

The gravitational acceleration, $\kappa_{h}$, on the acoustic black hole horizon surface, $r_{+}=r_{h}$, is given by
\begin{equation}
\kappa_{h} \equiv \frac{1}{2}\frac{1}{r_{h}^{2}}\left.\frac{d\Delta}{dr}\right|_{r=r_{h}}=\frac{1}{r_{h}}=\frac{c}{|A|}\ .
\label{eq:acel_grav_ext_draining_bathtub_Kerr}
\end{equation}
Then, an acoustic event horizon will emit an analogue Hawking radiation corresponding to a thermal bath of phonons at a temperature
\begin{equation}
T_{h}=\frac{\kappa_{h}}{2\pi}\ .
\label{eq:temp_Hawking_draining_bathtub_Kerr}
\end{equation}
%
%
\subsection{Analytic solutions of the massless Klein-Gordon equation}
In what follows we will consider the covariant Klein-Gordon equation, that describes the behavior of scalar fields in a curved spacetime, which has the form
\begin{equation}
\left[\frac{1}{\sqrt{-g}}\partial_{\rho}(g^{\rho\sigma}\sqrt{-g}\partial_{\sigma})\right]\Psi=0\ .
\label{eq:Klein-Gordon_cova_draining_bathtub_Kerr}
\end{equation}

Thus, the covariant Klein-Gordon equation in the spacetime of a rotating acoustic black hole given by the line element (\ref{eq:metrica_draining_bathtub_Kerr}), can be written as
\begin{eqnarray}
& & \left[-\frac{r^{3}}{\Delta}\frac{\partial^{2}}{\partial t^{2}}+\frac{\partial}{\partial r}\left(\frac{\Delta}{r}\frac{\partial}{\partial r}\right)+\left(\frac{1}{r}-\frac{B^{2}}{c^{2}r\Delta}\right)\frac{\partial^{2}}{\partial\phi^{2}}\right.\nonumber\\
& - & \left.\frac{2Br}{c\Delta}\frac{\partial^{2}}{\partial\phi\ \partial t}\right]\Psi(\mathbf{r},t)=0\ .
\label{eq:mov_draining_bathtub_Kerr}
\end{eqnarray}
According to the assumption that the spacetime under consideration is stationary \cite{PhysRevLett.46.1351} due to the fact that the fluid flow is assumed to be constant, the time dependence that solves Eq.~(\ref{eq:mov_draining_bathtub_Kerr}) may be separated as $\mbox{e}^{-i \omega t}$, where $\omega$ is the energy of the particles in the units chosen and we are considering that $\omega > 0$. Moreover, its axisymmetry permits us to separate the solution in $\phi$ as $\mbox{e}^{im\phi}$, where $m$ is a real constant that is not restricted to assume only a discrete set of values, because we are working with only two space dimensions. Thus, we can make the following ansatz
\begin{equation}
\Psi(\mathbf{r},t)=R(r)\mbox{e}^{im\phi}\mbox{e}^{-i \omega t}\ .
\label{eq:separacao_variaveis_draining_bathtub_Kerr}
\end{equation}
Substituting Eq.~(\ref{eq:separacao_variaveis_draining_bathtub_Kerr}) into (\ref{eq:mov_draining_bathtub_Kerr}), we find that
\begin{equation}
\frac{d}{dr}\left(\frac{\Delta}{r}\frac{dR}{dr}\right)+\left(\frac{\omega^{2}r^{3}}{\Delta}-\frac{m^{2}}{r}+\frac{m^{2}B^{2}}{c^{2}r\Delta}-\frac{2mB\omega r}{c\Delta}\right)R=0\ .
\label{eq:mov_radial_1_draining_bathtub_Kerr}
\end{equation}

Now, let us obtain the analytic and general solution for the radial part of Klein-Gordon equation given by Eq.~(\ref{eq:mov_radial_1_draining_bathtub_Kerr}). To do this, we will not take asymptotic limits, nor assume any condition on the frequency. On the contrary, we will consider the whole space, as well as the whole range of frequencies (or energies in the natural units), that is, $0 < \omega \leq \infty$. From our solution, we can get the low frequency regime solution obtained by Lepe and Saavedra \cite{PhysLettB.617.174} as a particular case.

Equation (\ref{eq:mov_radial_1_draining_bathtub_Kerr}) has an undesirable singularity at the origin, which hinders con\-sid\-er\-a\-bly its transformation to a Heun-type equation. However, introducing a new radial coordinate, $x$, such that
\begin{equation}
x=\frac{r^{2}}{2}\ ,
\label{eq:x_draining_bathtub_Kerr}
\end{equation}
and using this new coordinate, Eq.~(\ref{eq:Delta_draining_bathtub_Kerr}) turns into
\begin{equation}
\Delta=2x-\frac{A^{2}}{c^{2}}\ .
\label{eq:Delta_x_draining_bathtub_Kerr}
\end{equation}
Thus, from Eq.~(\ref{eq:Delta_x_draining_bathtub_Kerr}), we have that the new horizon surface of the rotating acoustic black hole is obtained from the condition
\begin{equation}
\Delta=2(x-x_{h})=0\ ,
\label{eq:superficie_hor_x_draining_bathtub_Kerr}
\end{equation}
where
\begin{equation}
x_{h}=\frac{1}{2}r_{h}^{2}
\label{eq:horizon_x_draining_bathtub_Kerr}
\end{equation}
is the root of $\Delta$ and corresponds to the new acoustic event horizon of the rotating acoustic black hole. Hence, using Eq.~(\ref{eq:superficie_hor_x_draining_bathtub_Kerr}), we can write down Eq.~(\ref{eq:mov_radial_1_draining_bathtub_Kerr}) as
\begin{eqnarray}
& & \frac{d^{2}R}{dx^{2}}+\left(\frac{1}{x-x_{h}}\right)\frac{dR}{dx}\nonumber\\
& + & \left[\frac{m^2 (B^2+2 c^2 x_{h})}{8 c^2 x_{h}^2}\frac{1}{x}+\frac{-B^2 m^2-2 c^2 m^2 x_{h}+4 c^2 x_{h}^2 \omega ^2}{8 c^2 x_{h}^2 }\frac{1}{x-x_{h}}\right.\nonumber\\
& + & \left.\frac{B^2 m^2-4 B c m x_{h} \omega +4 c^2 x_{h}^2 \omega ^2}{8 c^2 x_{h} }\frac{1}{(x-x_{h})^2}\right]R=0\ .
\label{eq:mov_radial_2_draining_bathtub_Kerr}
\end{eqnarray}

This equation has singularities at $x=(x_{1},x_{2})=(x_{h},0)$. The transformation of Eq.~(\ref{eq:mov_radial_2_draining_bathtub_Kerr}) to a Heun-type equation is achieved by setting the following homographic substitution
\begin{equation}
z=\frac{x-a_{1}}{a_{2}-a_{1}}=\frac{x-x_{h}}{0-x_{h}}\ .
\label{eq:z_draining_bathtub_Kerr}
\end{equation}
Thus, we can writte Eq.~(\ref{eq:mov_radial_2_draining_bathtub_Kerr}) as
\begin{eqnarray}
& & \frac{d^{2}R}{dz^{2}}+\frac{1}{z}\frac{dR}{dz}\nonumber\\
& + & \left\{\frac{B^2 m^2+2 c^2 m^2 x_{h}-4 c^2 x_{h}^2 \omega ^2}{8 c^2 x_{h}}\frac{1}{z}+\frac{-m^2 (B^2+2 c^2 x_{h})}{8 c^2 x_{h}}\frac{1}{z-1}\right.\nonumber\\
& - & \left[i\left(\frac{2 c x_{h} \omega -m B}{2 c \sqrt{2x_{h}}}\right)\right]^{2}\left.\frac{1}{z^2}\right\}R=0\ .
\label{eq:mov_radial_x_draining_bathtub_Kerr_3}
\end{eqnarray}
In what follows, let us perform an appropriate transformation in order to reduce the power of the term proportional to $1/z^{2}$ and put the coefficients that appear in the equation for the new function in an appropriate and convenient form \cite{Ronveaux:1995}. This transformation is the \textit{F-homotopic transformation} of the dependent variable $R(z) \mapsto U(z)$, such that
\begin{equation}
R(z)=z^{A_{1}}U(z)\ ,
\label{eq:F-homotopic}
\end{equation}
where the coefficient $A_{1}$ is given by
\begin{equation}
A_{1}=i\left(\frac{2 c x_{h} \omega -m B}{2 c \sqrt{2x_{h}}}\right)\ .
\label{eq:A1_draining_bathtub_Kerr_3}
\end{equation}
In this case, the function $U(z)$ satisfies the following equation
\begin{eqnarray}
& & \frac{d^{2}U}{dz^{2}}+\left(\frac{2A_{1}+1}{z}\right)\frac{dU}{dz}\nonumber\\
& + & \left[\frac{B^2 m^2+2 c^2 m^2 x_{h}-4 c^2 x_{h}^2 \omega ^2}{8 c^2 x_{h}}\frac{1}{z}-\frac{m^2 (B^2+2 c^2 x_{h})}{8 c^2 x_{h}}\frac{1}{z-1}\right]U=0\ ,
\label{eq:mov_radial_x_draining_bathtub_Kerr_4}
\end{eqnarray}
which is similar to the confluent Heun equation \cite{JPhysAMathTheor.43.035203}
\begin{equation}
\frac{d^{2}U}{dz^{2}}+\left(\alpha+\frac{\beta+1}{z}+\frac{\gamma+1}{z-1}\right)\frac{dU}{dz}+\left(\frac{\mu}{z}+\frac{\nu}{z-1}\right)U=0\ ,
\label{eq:Heun_confluente_forma_canonica}
\end{equation}
where $U(z)=\mbox{HeunC}(\alpha,\beta,\gamma,\delta,\eta;z)$ are the confluent Heun functions, with the parameters $\alpha$, $\beta$, $\gamma$, $\delta$ and $\eta$, related to $\mu$ and $\nu$ by
\begin{equation}
\mu=\frac{1}{2}(\alpha-\beta-\gamma+\alpha\beta-\beta\gamma)-\eta\ ,
\label{eq:mu_Heun_conlfuente_2}
\end{equation}
\begin{equation}
\nu=\frac{1}{2}(\alpha+\beta+\gamma+\alpha\gamma+\beta\gamma)+\delta+\eta\ ,
\label{eq:nu_Heun_conlfuente_2}
\end{equation}
according to the standard package of the \textbf{Maple}\texttrademark \textbf{17}.

Thus, the general solution of the radial part of the Klein-Gordon equation for a massless scalar field in the spacetime of a rotating acoustic black hole, in the region exterior to the acoustic event horizon, given by Eq.~(\ref{eq:mov_radial_x_draining_bathtub_Kerr_3}), over the entire range $0 \leq z < \infty$, can be written, following the details presented in refs. \cite{Ronveaux:1995,JPhysAMathTheor.43.035203}, as
\begin{eqnarray}
R(z) & = & z^{\frac{\beta}{2}}\nonumber\\
& \times & \{C_{1}\ \mbox{HeunC}(\alpha,\beta,\gamma,\delta,\eta;z)+C_{2}\ z^{-\beta}\ \mbox{HeunC}(\alpha,-\beta,\gamma,\delta,\eta;z)\}\ ,\nonumber\\
\label{eq:solucao_geral_radial_draining_bathtub_Kerr}
\end{eqnarray}
where $C_{1}$ and $C_{2}$ are constants, and the parameters $\alpha$, $\beta$, $\gamma$, $\delta$, and $\eta$ are now given by:
\begin{equation}
\alpha=0\ ;
\label{eq:alpha_radial_HeunC_draining_bathtub_Kerr}
\end{equation}
\begin{equation}
\beta=i\left(\frac{2 c x_{h} \omega -m B}{c \sqrt{2x_{h}}}\right)\ ;
\label{eq:beta_radial_HeunC_draining_bathtub_Kerr}
\end{equation}
\begin{equation}
\gamma=-1\ ;
\label{eq:gamma_radial_HeunC_draining_bathtub_Kerr}
\end{equation}
\begin{equation}
\delta=-\frac{x_{h} \omega ^2}{2}\ ;
\label{eq:delta_radial_HeunC_draining_bathtub_Kerr}
\end{equation}
\begin{equation}
\eta=\frac{1}{8} \left[m^2 \left(-\frac{B^2}{c^2 x_{h}}-2\right)+4 x_{h} \omega ^2+4\right]\ .
\label{eq:eta_radial_HeunC_draining_bathtub_Kerr}
\end{equation}
These two functions form linearly independent solutions of the confluent Heun dif\-fer\-en\-tial equation due to the fact that $\beta$ is not necessarily an integer. 

In what follows we will consider the expansion of the solutions in Eq.~(\ref{eq:solucao_geral_radial_draining_bathtub_Kerr}), which is regular at $z=0$ and appropriate to study the problem concerning the analogue Hawking radiation. The expansion in power series of the confluent Heun functions with respect to the independent variable $z$, in a neighborhood of the regular singular point $z=0$ \cite{Ronveaux:1995}, can writte as
\begin{eqnarray}
\mbox{HeunC}(\alpha,\beta,\gamma,\delta,\eta;z) & = & 1+\frac{1}{2}\frac{(-\alpha\beta+\beta\gamma+2\eta-\alpha+\beta+\gamma)}{(\beta+1)}z\nonumber\\
& + & \frac{1}{8}\frac{1}{(\beta+1)(\beta+2)}(\alpha^{2}\beta^{2}-2\alpha\beta^{2}\gamma+\beta^{2}\gamma^{2}\nonumber\\
& - & 4\eta\alpha\beta+4\eta\beta\gamma+4\alpha^{2}\beta-2\alpha\beta^{2}-6\alpha\beta\gamma\nonumber\\
& + & 4\beta^{2}\gamma+4\beta\gamma^{2}+4\eta^{2}-8\eta\alpha+8\eta\beta+8\eta\gamma\nonumber\\
& + & 3\alpha^{2}-4\alpha\beta-4\alpha\gamma+3\beta^{2}+4\beta\delta\nonumber\\
& + & 10\beta\gamma+3\gamma^{2}+8\eta+4\beta+4\delta+4\gamma)z^2+...\ .
\label{eq:serie_HeunC_todo_z}
\end{eqnarray}
%
%
\subsection{Analogue Hawking radiation}\label{Hawking_radiation_rotating}
We will consider the massless scalar field near the acoustic horizon in order to discuss the analogue Hawking radiation. From Eqs.~(\ref{eq:x_draining_bathtub_Kerr}), (\ref{eq:horizon_x_draining_bathtub_Kerr}), (\ref{eq:z_draining_bathtub_Kerr}) and (\ref{eq:serie_HeunC_todo_z}), we can see that the radial solution given by Eq.~(\ref{eq:solucao_geral_radial_draining_bathtub_Kerr}), near the acoustic event horizon, that is, when $r \rightarrow r_{h} \Rightarrow x \rightarrow x_{h} \Rightarrow z \rightarrow 0$, behaves asymptotically as:
\begin{equation}
R(z) \sim z^{\frac{\beta}{2}}\{C_{1} \cdot 1 + C_{2} \cdot z^{-\beta} \cdot 1\}=C_{1} z^{\frac{\beta}{2}} + C_{2} z^{-\frac{\beta}{2}}\ ,
\label{eq:exp_0_solucao_geral_radial_draining_bathtub_Kerr_1}
\end{equation}
or
\begin{equation}
R(r) \sim \frac{C_{1}}{-r_{h}^{\frac{\beta}{2}}}(r-r_{h})^{\frac{\beta}{2}} + \frac{C_{2}}{-r_{h}^{-\frac{\beta}{2}}}(r-r_{h})^{\frac{-\beta}{2}}\ .
\label{eq:exp_0_solucao_geral_radial_draining_bathtub_Kerr_2}
\end{equation}
Therefore, we can write
\begin{equation}
R(r) \sim C_{1} (r-r_{h})^{\frac{\beta}{2}}+C_{2} (r-r_{h})^{-\frac{\beta}{2}}\ ,
\label{eq:exp_0_solucao_geral_radial_draining_bathtub_Kerr}
\end{equation}
where we are considering contributions only of the first term in the expansion, and all constants are included in $C_{1}$ and $C_{2}$. Thus, taking into account the solution of the time dependence, near the rotating acoustic black hole event horizon $r_{h}$, we can write
\begin{equation}
\Psi=\mbox{e}^{-i \omega t}(r-r_{h})^{\pm\frac{\beta}{2}}\ .
\label{eq:sol_onda_radial_draining_bathtub_Kerr}
\end{equation}
From Eq.~(\ref{eq:beta_radial_HeunC_draining_bathtub_Kerr}), for the parameter $\beta$, we obtain
\begin{equation}
\beta=ir_{h}\left(\omega-m\frac{B}{cr_{h}^{2}}\right)\ .
\label{eq:beta/2_solucao_geral_radial_draining_bathtub_Kerr}
\end{equation}
Then, substituting Eqs.~(\ref{eq:horizon}) and (\ref{eq:acel_grav_ext_draining_bathtub_Kerr}) into (\ref{eq:beta/2_solucao_geral_radial_draining_bathtub_Kerr}), we get
\begin{equation}
\beta=\frac{i}{\kappa_{h}}(\omega-\omega_{h})\ ,
\label{eq:expoente_rad_Hawking_draining_bathtub_Kerr}
\end{equation}
where 
\begin{equation}
\omega_{h}=m\Omega_{h}\ ,
\label{eq:omega0}
\end{equation}
with
\begin{equation}
\Omega_{h}=\frac{Bc}{A^{2}}
\label{eq:vel_ang_draining_bathtub_Kerr}
\end{equation}
being the dragging angular velocity of the acoustic event horizon.

Therefore, on the rotating acoustic black hole horizon surface, the ingoing and outgoing wave solutions are
\begin{equation}
\Psi_{in}=\mbox{e}^{-i \omega t}(r-r_{h})^{-\frac{i}{2\kappa_{h}}(\omega-\omega_{h})}\ ,
\label{eq:sol_in_1_draining_bathtub_Kerr}
\end{equation}
\begin{equation}
\Psi_{out}(r>r_{h})=\mbox{e}^{-i \omega t}(r-r_{h})^{\frac{i}{2\kappa_{h}}(\omega-\omega_{h})}\ .
\label{eq:sol_out_2_draining_bathtub_Kerr}
\end{equation}
These solutions for the scalar fields near the acoustic horizon will be useful to investigate the Hawking radiation of massless scalar particles. It is worth calling attention to the fact that we are using the analytic solution of the radial part of Klein-Gordon equation in the spacetime under consideration, differently from the calculations usually done in the literature \cite{ClassQuantumGrav.20.3907,PhysLettB.617.174}.

Using the definitions of the tortoise and Eddington-Finkelstein coordinates, given by \cite{JHighEnergyPhys.07.070}
\begin{equation}
dr_{*}=\frac{r^{2}}{\Delta}dr\ ,
\label{eq:coord_tortoise_1.1}
\end{equation}
we have
\begin{equation}
\ln(r-r_{h})=\frac{1}{r_{h}^{2}}\left.\frac{d\Delta}{dr}\right|_{r=r_{h}}r_{*}=2\kappa_{h}r_{*}\ ,
\label{eq:coord_tortoise_1}
\end{equation}
\begin{equation}
\hat{r}=\frac{\omega-\omega_{h}}{\omega}r_{*}\ ,
\label{eq:hatr}
\end{equation}
\begin{equation}
v=t+\hat{r}
\label{eq:coord_Eddington-Finkelstein}\ ,
\end{equation}
and thus the following ingoing wave solution can be obtained
\begin{eqnarray}
\Psi_{in} & = & \mbox{e}^{-i \omega v}\mbox{e}^{i \omega \hat{r}}(r-r_{h})^{-\frac{i}{2\kappa_{h}}(\omega-\omega_{h})}\nonumber\\
& = & \mbox{e}^{-i \omega v}\mbox{e}^{i (\omega-\omega_{h}) r_{*}}(r-r_{h})^{-\frac{i}{2\kappa_{h}}(\omega-\omega_{h})}\nonumber\\
& = & \mbox{e}^{-i \omega v}(r-r_{h})^{\frac{i}{2\kappa_{h}}(\omega-\omega_{h})}(r-r_{h})^{-\frac{i}{2\kappa_{h}}(\omega-\omega_{h})}\nonumber\\
& = & \mbox{e}^{-i \omega v}\ .
\label{eq:sol_in_1_draining_bathtub_Kerr_tortoise}
\end{eqnarray}
The outgoing wave solution is given by
\begin{eqnarray}
\Psi_{out}(r>r_{h}) & = & \mbox{e}^{-i \omega v}\mbox{e}^{i \omega \hat{r}}(r-r_{h})^{\frac{i}{2\kappa_{h}}(\omega-\omega_{h})}\nonumber\\
& = & \mbox{e}^{-i \omega v}\mbox{e}^{i (\omega-\omega_{h}) r_{*}}(r-r_{h})^{\frac{i}{2\kappa_{h}}(\omega-\omega_{h})}\nonumber\\
& = & \mbox{e}^{-i \omega v}(r-r_{h})^{\frac{i}{2\kappa_{h}}(\omega-\omega_{h})}(r-r_{h})^{\frac{i}{2\kappa_{h}}(\omega-\omega_{h})}\nonumber\\
& = & \mbox{e}^{-i \omega v}(r-r_{h})^{\frac{i}{\kappa_{h}}(\omega-\omega_{h})}\ .
\label{eq:sol_out_2_draining_bathtub_Kerr_tortoise}
\end{eqnarray}

Note that, if we put $Q=0$ into Eqs.~(96) and (97) of Ref. \cite{AnnPhys.350.14}, the resulting solutions are analyticly analogous to solutions above given by Eqs.~(\ref{eq:sol_in_1_draining_bathtub_Kerr_tortoise}) and (\ref{eq:sol_out_2_draining_bathtub_Kerr_tortoise}).
%
%
\subsection{Analytic extension and radiation spectrum}
Now, let us obtain by analytic continuation a real damped part of the outgoing wave solution of the massless scalar field which will be used to construct an explicit expression for the decay rate $\Gamma_{h}$. This real damped part corresponds (at least partially) to the temporal contribution to the decay rate \cite{PhysLettB.698.438} found by the tunneling method used to investigate the analogue Hawking radiation.

From Eq.~(\ref{eq:sol_out_2_draining_bathtub_Kerr_tortoise}), we see that this solution is not analytical in the acoustic event horizon $r=r_{h}$. By analytic continuation, rotating by an angle $-\pi$ through the lower-half complex $r$ plane, we obtain
\begin{equation}
(r-r_{h}) \rightarrow \left|r-r_{h}\right|\mbox{e}^{-i\pi}=(r_{h}-r)\mbox{e}^{-i\pi}\ .
\label{eq:rel_3_draining_bathtub_Kerr}
\end{equation}
Thus, the outgoing wave solution on the acoustic horizon surface $r_{h}$ is
\begin{equation}
\Psi_{out}(r<r_{h})=\mbox{e}^{-i\omega v}(r_{h}-r)^{\frac{i}{\kappa_{h}}(\omega-\omega_{h})}\mbox{e}^{\frac{\pi}{\kappa_{h}}(\omega-\omega_{h})}\ .
\label{eq:sol_1_out_4_draining_bathtub_Kerr}
\end{equation}
Equations (\ref{eq:sol_out_2_draining_bathtub_Kerr_tortoise}) and (\ref{eq:sol_1_out_4_draining_bathtub_Kerr}) describe the outging wave outside and inside of the rotating acoustic black hole event horizon, respectively. Therefore, for an outgoing wave of a particle with energy $\omega > 0$, the outgoing decay rate or the relative scattering probability of the scalar wave at the acoustic event horizon surface, $r=r_{h}$, is given by
\begin{equation}
\Gamma_{h}=\left|\frac{\Psi_{out}(r>r_{h})}{\Psi_{out}(r<r_{h})}\right|^{2}=\mbox{e}^{-\frac{2\pi}{\kappa_{h}}(\omega-\omega_{h})}\ .
\label{eq:taxa_refl_draining_bathtub_Kerr}
\end{equation}
This result was already formally obtained in the literature \cite{PhysLettB.698.438}, in different context, and is analogous to the one obtained in \cite{AnnPhys.350.14}, for an astrophysical black hole.

According to the Damour-Ruffini-Sannan method \cite{PhysRevD.14.332,GenRelativGravit.20.239} for astrophysical black holes, a correct wave describing a particle flying off of the rotating acoustic black hole is given by
\begin{eqnarray}
\Psi_{\omega}(r) & = & N_{\omega}\ [\ H(r-r_{h})\ \Psi_{\omega}^{out}(r-r_{h})\nonumber\\
& + & H(r_{h}-r)\ \Psi_{\omega}^{out}(r_{h}-r)\ \mbox{e}^{\frac{\pi}{\kappa_{h}}(\omega-\omega_{h})}\ ]\ ,
\label{eq:solucao_geral_onda_out_draining_bathtub_Kerr}
\end{eqnarray}
where $N_{\omega}$ is the normalization constant, such that
\begin{equation}
\left\langle \Psi_{\omega_{1}}(r) | \Psi_{\omega_{2}}(r) \right\rangle=-\delta(\omega_{1}-\omega_{2})\ ,
\label{eq:cond_norm}
\end{equation}
with $H(x)$ being the Heaviside function and $\Psi_{\omega}^{out}(x)$ the normalized wave functions given, from Eq.~(\ref{eq:sol_out_2_draining_bathtub_Kerr_tortoise}), by
\begin{equation}
\Psi_{\omega}^{out}(x)=\mbox{e}^{-i \omega v}x^{\frac{i}{\kappa_{h}}(\omega-\omega_{h})}\ .
\label{eq:sol_out_2_draining_bathtub_Kerr_tortoise_x}
\end{equation}

Thus, from the normalization condition
\begin{equation}
\left\langle \Psi_{\omega}(r) | \Psi_{\omega}(r) \right\rangle=1=\left|N_{\omega}\right|^{2}\left[\mbox{e}^{\frac{2\pi}{\kappa_{h}}(\omega-\omega_{h})}-1\right]\ ,
\label{eq:norm_onda_out_draining_bathtub_Kerr}
\end{equation}
we get the resulting analogue Hawking radiation spectrum of scalar particles, which is given by
\begin{equation}
\left|N_{\omega}\right|^{2}=\frac{1}{\mbox{e}^{\frac{2\pi}{\kappa_{h}}(\omega-\omega_{h})}-1}=\frac{1}{\mbox{e}^{\frac{\hbar(\omega-\omega_{h})}{k_{B}T_{h}}}-1}\ ,
\label{eq:espectro_rad_draining_bathtub_Kerr_2}
\end{equation}
where the Boltzmann's and Planck's constants were reintroduced.

Therefore, we can see that the resulting analogue Hawking radiation spectrum of massless scalar particles has a thermal character, analogous to the black body spectrum, where $k_{B}T_{h}=\hbar\kappa_{h}/2\pi$.
%
%
\section{Canonical acoustic black hole}
A solution for a spherically symmetric flow of incompressible fluid, called canonical acoustic black hole, was found by Visser in \cite{ClassQuantumGrav.15.1767}. The acoustic metric which appropriatelly describes this situation is given by
\begin{equation}
ds^{2}=-c^{2}\ dt^{2}+\left(dr \pm c\frac{r_{0}^{2}}{r^{2}}\ dt\right)^{2}+r^{2}(d\theta^{2}+\sin^{2}\theta\ d\phi^{2})\ ,
\label{eq:metrica_canonical}
\end{equation}
where $c$ is the speed of sound and is constant throughout the fluid flow, and is defined in terms of the velocity $v$ as
\begin{equation}
v=c\frac{r_{0}^{2}}{r^{2}}\ .
\label{eq:velocidade_canonical}
\end{equation}
Using the Schwarzschild time coordinate $\tau$ instead of the laboratory time $t$, and doing the coordinate transformation \cite{ClassQuantumGrav.15.1767}
\begin{equation}
d\tau=dt \pm \frac{r_{0}^{2}/r^{2}}{c[1-(r_{0}^{4}/r^{4})]}\ dr\ ,
\label{eq:relacao1_canonical}
\end{equation}
the line element that describes a canonical acoustic black hole can be rewritten as
\begin{equation}
ds^{2}=-\frac{c^{2}}{r^{4}}\Delta\ d\tau^{2}+\frac{r^{4}}{\Delta}\ dr^{2}+r^{2}(d\theta^{2}+\sin^{2}\theta\ d\phi^{2})\ ,
\label{eq:metrica_canonical_Schw}
\end{equation}
where
\begin{equation}
\Delta=r^{4}-r_{0}^{4}\ .
\label{eq:Delta_canonical_Schw}
\end{equation}
Obviously, the form of this metric is different from that of the standard geometries typically considered in general relativity.

From Eq.~(\ref{eq:metrica_canonical_Schw}) we conclude that the radius of the acoustic event horizon is given by
\begin{equation}
r_{h}=r_{0}\ ,
\label{eq:horizon_canonical_Schw}
\end{equation}
where $r_{0}$ is obtained from Eq.~(\ref{eq:velocidade_canonical}), and corresponds to the event horizon of the canonical acoustic black hole.

The gravitational acceleration, $\kappa_{0}$, on the acoustic black hole horizon surface, $r_{0}$, is given by \cite{arXiv:9901047}
\begin{equation}
\kappa_{0} \equiv \left|\left.\frac{\partial v^{r}}{\partial r}\right|_{r=r_{0}}\right|=\frac{2c}{r_{0}}\ .
\label{eq:acel_grav_ext_canonical_Schw}
\end{equation}
Then, an acoustic event horizon will emit analogue Hawking radiation in the form of a thermal bath of phonons at a temperature
\begin{equation}
T_{0}=\frac{\kappa_{0}}{2\pi c}=\frac{1}{\pi r_{0}}\ .
\label{eq:temp_Hawking_canonical_Schw}
\end{equation}
%
%
\subsection{Analytic solutions of the massless Klein-Gordon equation}
The covariant Klein-Gordon equation in the spacetime of a canonical acoustic black hole given by the line element (\ref{eq:metrica_canonical_Schw}), can be written as
\begin{eqnarray}
& & \left[-\frac{r^{6}}{c^{2}\Delta}\frac{\partial^{2}}{\partial \tau^{2}}+\frac{\partial}{\partial r}\left(\frac{\Delta}{r^{2}}\frac{\partial}{\partial r}\right)\right.\nonumber\\
& + & \left.\frac{1}{\sin\theta}\frac{\partial}{\partial\theta}\left(\sin\theta\frac{\partial}{\partial\theta}\right)+\frac{1}{\sin^{2}\theta}\frac{\partial^{2}}{\partial\phi^{2}}\right]\Psi(\mathbf{r},\tau)=0\ .
\label{eq:mov_canonical_Schw}
\end{eqnarray}
The spacetime under consideration is static, so the time dependence that solves Eq.~(\ref{eq:mov_canonical_Schw}) may be separated as $\mbox{e}^{-i \omega \tau}$, where $\omega$ is the energy of the particles in the units chosen. Moreover, rotational invariance with respect to $\phi$ implies that the solution in $\phi$ is $\mbox{e}^{im\phi}$, where $m=\pm 1,\pm 2, \pm 3,...$, is the azimuthal quantum number. Thus, the general angular solution is given in terms of the spherical harmonic function $Y_{m}^{l}(\theta,\phi)=P_{m}^{l}(\cos\theta)\mbox{e}^{im\phi}$, where $l$ is an integer such that $|m| \leq l$ \cite{Arfken:2005}. Therefore, $\Psi(\mathbf{r},\tau)$ can be written as
\begin{equation}
\Psi(\mathbf{r},\tau)=R(r)Y_{m}^{l}(\theta,\phi)\mbox{e}^{-i \omega \tau}\ .
\label{eq:separacao_variaveis_canonical_Schw}
\end{equation}
Substituting Eq.~(\ref{eq:separacao_variaveis_canonical_Schw}) into (\ref{eq:mov_canonical_Schw}), we find that
\begin{equation}
\frac{d}{dr}\left(\frac{\Delta}{r^{2}}\frac{dR}{dr}\right)+\left(\frac{\omega^{2}r^{6}}{c^{2}\Delta}-\lambda_{lm}\right)R=0\ ,
\label{eq:mov_radial_1_canonical_Schw}
\end{equation}
where $\lambda_{lm}=l(l+1)$.

Now, let us obtain the analytic and general solution for the radial part of Klein-Gordon equation given by Eq.~(\ref{eq:mov_radial_1_canonical_Schw}). This equation has four singularities, which hinders considerably its transformation to a Heun-type equation. Then, we define a new radial coordinate, $x$, such that
\begin{equation}
x=\frac{r^{2}}{2}\ .
\label{eq:x_canonical_Schw}
\end{equation}
Using this new coordinate, Eq.~(\ref{eq:Delta_canonical_Schw}) can be rewritten as
\begin{equation}
\Delta=4(x^{2}-x_{0})\ ,
\label{eq:Delta_x_canonical_Schw}
\end{equation}
where
\begin{equation}
x_{0}=\frac{1}{4}r_{0}^{4}\ .
\label{eq:horizon_canonical_Schw_x_canonical_Schw}
\end{equation}
Thus, from Eq.~(\ref{eq:Delta_x_canonical_Schw}), we have that the new acoustic horizon surface equation of the canonical acoustic black hole is obtained from the condition
\begin{equation}
\Delta=4(x-x_{+})(x-x_{-})=0\ ,
\label{eq:superficie_hor_x_canonical_Schw}
\end{equation}
whose solutions are
\begin{equation}
x_{+}=x_{0}\ ,
\label{eq:sol_padrao_Lense-Thirring_1}
\end{equation}
\begin{equation}
x_{-}=-x_{0}\ ,
\label{eq:sol_padrao_Lense-Thirring_2}
\end{equation}
and correspond to event horizons of the canonical acoustic black hole. Hence, using Eq.~(\ref{eq:superficie_hor_x_canonical_Schw}), we can write down Eq.~(\ref{eq:mov_radial_1_canonical_Schw}) as
\begin{eqnarray}
& & \frac{d^{2}R}{dx^{2}}+\left(\frac{1}{x-x_{+}}+\frac{1}{x-x_{-}}\right)\frac{dR}{dx}\nonumber\\
& + & \frac{1}{(x-x_{+})(x-x_{-})}\left[\frac{\omega ^2}{2 c^2}x-\frac{c^2 \lambda_{lm} -2 x_{+} \omega ^2-2 x_{-} \omega ^2}{4 c^2}\right.\nonumber\\
& + & \left.\frac{x_{+}^3 \omega ^2}{2 c^2  (x_{+}-x_{-})}\frac{1}{x-x_{+}}-\frac{x_{-}^3 \omega ^2}{2 c^2  (x_{+}-x_{-})}\frac{1}{x-x_{-}}\right]R=0\ .
\label{eq:mov_radial_2_canonical_Schw}
\end{eqnarray}

This equation has singularities at $x=(a_{1},a_{2})=(x_{+},x_{-})$, and at $x=\infty$. The transformation of Eq.~(\ref{eq:mov_radial_2_canonical_Schw}) to a Heun-type equation is achieved by setting the following homographic substitution
\begin{equation}
z=\frac{x-a_{1}}{a_{2}-a_{1}}=\frac{x-x_{+}}{x_{-}-x_{+}}\ .
\label{eq:z_canonical_Schw}
\end{equation}
Thus, we can writte Eq.~(\ref{eq:mov_radial_2_canonical_Schw}) as
\begin{eqnarray}
& & \frac{d^{2}R}{dz^{2}}+\left(\frac{1}{z}+\frac{1}{z-1}\right)\frac{dR}{dz}\nonumber\\
& + & \left\{\frac{c^2 \lambda_{lm}  x_{+}^2-2 c^2 \lambda_{lm}  x_{+} x_{-}+c^2 \lambda_{lm}  x_{-}^2-2 x_{+}^3 \omega ^2+6 x_{+}^2 x_{-} \omega ^2}{4 c^2 (x_{+}-x_{-})^2}\frac{1}{z}\right.\nonumber\\
& + & \frac{-c^2 \lambda_{lm}  x_{+}^2+2 c^2 \lambda_{lm}  x_{+} x_{-}-c^2 \lambda_{lm}  x_{-}^2-6 x_{+} x_{-}^2 \omega ^2+2 x_{-}^3 \omega ^2}{4 c^2 (x_{+}-x_{-})^2}\frac{1}{z-1}\nonumber\\
& - & \left[i\frac{x_{+}^{3/2} \omega}{\sqrt{2} c  (x_{+}-x_{-})}\right]^{2}\frac{1}{z^2}+\left[i\frac{x_{-}^{3/2} \omega}{\sqrt{2} c  (x_{+}-x_{-})}\right]^{2}\left.\frac{1}{(z-1)^2}\right\}R=0\ .
\label{eq:mov_radial_x_canonical_Schw_3}
\end{eqnarray}
Now, let us perform a transformation in order to reduce the power of the terms proportional to $1/z^{2}$ and $1/(z-1)^{2}$ and write the coefficients that appear in the equation for the new function in a useful form \cite{Ronveaux:1995}. This transformation is the \textit{F-homotopic trans\-for\-ma\-tion} of the dependent variable $R(z) \mapsto U(z)$, such that
\begin{equation}
R(z)=z^{A_{1}}(z-1)^{A_{2}}U(z)\ ,
\label{eq:F-homotopic_canonical_Schw}
\end{equation}
where the coefficients $A_{1}$ and $A_{2}$ are given by
\begin{equation}
A_{1}=i\frac{x_{+}^{3/2} \omega}{\sqrt{2} c  (x_{+}-x_{-})}\ ,
\label{eq:A1_canonical_Schw_3}
\end{equation}
\begin{equation}
A_{2}=i\frac{x_{-}^{3/2} \omega}{\sqrt{2} c  (x_{+}-x_{-})}\ .
\label{eq:A2_canonical_Schw_3}
\end{equation}
In this case, the function $U(z)$ satisfies the following equation
\begin{eqnarray}
& & \frac{d^{2}U}{dz^{2}}+\left(\frac{2A_{1}+1}{z}+\frac{2A_{2}+1}{z-1}\right)\frac{dU}{dz}\nonumber\\
& + & \left[\frac{-A_{1}-A_{2}-2A_{1}A_{2}+A_{3}}{z}+\frac{A_{1}+A_{2}+2A_{1}A_{2}+A_{4}}{z-1}\right]U=0\ ,
\label{eq:mov_radial_x_canonical_Schw_4}
\end{eqnarray}
where the coefficients $A_{3}$ and $A_{4}$ are given by
\begin{equation}
A_{3}=\frac{c^2 \lambda_{lm}  x_{+}^2-2 c^2 \lambda_{lm}  x_{+} x_{-}+c^2 \lambda_{lm}  x_{-}^2-2 x_{+}^3 \omega ^2+6 x_{+}^2 x_{-} \omega ^2}{4 c^2 (x_{+}-x_{-})^2}\ ,
\label{eq:A3_canonical_Schw_3}
\end{equation}
\begin{equation}
A_{4}=\frac{-c^2 \lambda_{lm}  x_{+}^2+2 c^2 \lambda_{lm}  x_{+} x_{-}-c^2 \lambda_{lm}  x_{-}^2-6 x_{+} x_{-}^2 \omega ^2+2 x_{-}^3 \omega ^2}{4 c^2 (x_{+}-x_{-})^2}\ .
\label{eq:A4_canonical_Schw_3}
\end{equation}
Note that Eq.~(\ref{eq:mov_radial_x_canonical_Schw_4}) is similar to the confluent Heun equation (\ref{eq:Heun_confluente_forma_canonica}). Thus, the general solution of the radial part of the Klein-Gordon equation for a massless scalar particle in the spacetime under consideration, in the exterior region of the acoustic event horizon, given by Eq.~(\ref{eq:mov_radial_x_canonical_Schw_3}) over the entire range $0 \leq z < \infty$, can be written as \cite{Ronveaux:1995,JPhysAMathTheor.43.035203}
\begin{eqnarray}
R(z) & = & z^{\frac{1}{2}\beta}(z-1)^{\frac{1}{2}\gamma}\nonumber\\
& \times & \{C_{1}\ \mbox{HeunC}(\alpha,\beta,\gamma,\delta,\eta;z)+C_{2}\ z^{-\beta}\ \mbox{HeunC}(\alpha,-\beta,\gamma,\delta,\eta;z)\}\ ,\nonumber\\
\label{eq:solucao_geral_radial_canonical_Schw}
\end{eqnarray}
where $C_{1}$ and $C_{2}$ are constants, and the parameters $\alpha$, $\beta$, $\gamma$, $\delta$, and $\eta$ are now given by:
\begin{equation}
\alpha=0\ ;
\label{eq:alpha_radial_HeunC_canonical_Schw}
\end{equation}
\begin{equation}
\beta=i \sqrt{2} \frac{x_{+}^{3/2} \omega}{ c  (x_{+}-x_{-})}\ ;
\label{eq:beta_radial_HeunC_canonical_Schw}
\end{equation}
\begin{equation}
\gamma=i \sqrt{2} \frac{x_{-}^{3/2} \omega}{ c  (x_{+}-x_{-})}\ ;
\label{eq:gamma_radial_HeunC_canonical_Schw}
\end{equation}
\begin{equation}
\delta=-\frac{\omega ^2 (x_{+}-x_{-})}{2 c^2}\ ;
\label{eq:delta_radial_HeunC_canonical_Schw}
\end{equation}
\begin{equation}
\eta=\frac{2 x_{+}^2 \omega ^2 (x_{+}-3 x_{-})-c^2 \lambda  (x_{+}-x_{-})^2}{4 c^2 (x_{+}-x_{-})^2}\ .
\label{eq:eta_radial_HeunC_canonical_Schw}
\end{equation}
These two functions form linearly independent solutions of the confluent Heun dif\-fer\-en\-tial equation provided $\beta$ is not integer, which is a condition satisfied by this parameter because there is no physical reason to impose that $\beta$ should be an integer.
%
%
\subsection{Analogue Hawking radiation}\label{Hawking_radiation_canonical}
Now, let us consider the massless scalar field near the horizon in order to discuss the analogue Hawking radiation. To do this, from Eqs.~(\ref{eq:z_canonical_Schw}) and (\ref{eq:serie_HeunC_todo_z}), and following the same procedure of subsection (\ref{Hawking_radiation_rotating}), we can see that the radial solution given by Eq.~(\ref{eq:solucao_geral_radial_canonical_Schw}), near the acoustic event horizon, that is, when $r \rightarrow r_{0} \Rightarrow x \rightarrow x_{+} \Rightarrow z \rightarrow 0$, behaves asymptotically as
\begin{equation}
R(r) \sim C_{1}\ (r-r_{0})^{\beta/2}+C_{2}\ (r-r_{0})^{-\beta/2}\ ,
\label{eq:exp_0_solucao_geral_radial_canonical_Schw}
\end{equation}
where we are considering contributions arising only from the first term in the expansion, and all constants are included in $C_{1}$ and $C_{2}$. Thus, taking into account the time dependence of the solution, near the canonical acoustic black hole event horizon $r_{0}$, we have
\begin{equation}
\Psi=\mbox{e}^{-i \omega \tau}(r-r_{0})^{\pm\beta/2}\ .
\label{eq:sol_onda_radial_canonical_Schw}
\end{equation}
From Eq.~(\ref{eq:beta_radial_HeunC_canonical_Schw}), for the parameter $\beta$, we obtain
\begin{equation}
\beta=i\frac{r_{0}}{2c}\omega\ .
\label{eq:beta/2_solucao_geral_radial_canonical_Schw}
\end{equation}
Then, substituting Eqs.~(\ref{eq:horizon_canonical_Schw}) and (\ref{eq:acel_grav_ext_canonical_Schw}) into (\ref{eq:beta/2_solucao_geral_radial_canonical_Schw}), we get
\begin{equation}
\beta=\frac{i}{\kappa_{0}}\omega\ .
\label{eq:expoente_rad_Hawking_canonical_Schw}
\end{equation}

Therefore, on the canonical acoustic black hole horizon surface, the ingoing and outgoing wave solutions are
\begin{equation}
\Psi_{in}=\mbox{e}^{-i \omega \tau}(r-r_{0})^{-\frac{i}{2\kappa_{0}}\omega}\ ,
\label{eq:sol_in_1_canonical_Schw}
\end{equation}
\begin{equation}
\Psi_{out}(r>r_{0})=\mbox{e}^{-i \omega \tau}(r-r_{0})^{\frac{i}{2\kappa_{0}}\omega}\ .
\label{eq:sol_out_2_canonical_Schw}
\end{equation}
These solutions for the scalar fields near the acoustic horizon obtained from the analytical solution of the radial part of the Klein-Gordon equation in the background under consideration will be useful to investigate the analogue Hawking radiation.

Using the definitions of the tortoise and Eddington-Finkelstein coordinates, given by
\begin{equation}
dr_{*}=\frac{r^{4}}{c}\frac{1}{\Delta}dr\ ,
\label{eq:coord_tortoise_1.1_canonical_Schw}
\end{equation}
we have
\begin{equation}
\ln(r-r_{0})=\frac{c}{r_{0}^{4}}\left.\frac{d\Delta}{dr}\right|_{r=r_{0}}r_{*}=2\kappa_{0}r_{*}\ ,
\label{eq:coord_tortoise_1_canonical_Schw}
\end{equation}
\begin{equation}
\hat{r}=\frac{\omega-\omega_{0}}{\omega}r_{*}\ ,
\label{eq:hatr_canonical_Schw}
\end{equation}
\begin{equation}
v=\tau+\hat{r}
\label{eq:coord_Eddington-Finkelstein_canonical_Schw}\ ,
\end{equation}
and thus the following ingoing wave solution can be obtained
\begin{eqnarray}
\Psi_{in} & = & \mbox{e}^{-i \omega v}\mbox{e}^{i \omega \hat{r}}(r-r_{0})^{-\frac{i}{2\kappa_{0}}\omega}\nonumber\\
& = & \mbox{e}^{-i \omega v}\mbox{e}^{i \omega r_{*}}(r-r_{0})^{-\frac{i}{2\kappa_{0}}\omega}\nonumber\\
& = & \mbox{e}^{-i \omega v}(r-r_{0})^{\frac{i}{2\kappa_{0}}\omega}(r-r_{0})^{-\frac{i}{2\kappa_{0}}\omega}\nonumber\\
& = & \mbox{e}^{-i \omega v}\ .
\label{eq:sol_in_1_canonical_Schw_tortoise}
\end{eqnarray}
Otherwise, the outgoing wave solution is given by
\begin{eqnarray}
\Psi_{out}(r>r_{0}) & = & \mbox{e}^{-i \omega v}\mbox{e}^{i \omega \hat{r}}(r-r_{0})^{\frac{i}{2\kappa_{0}}\omega}\nonumber\\
& = & \mbox{e}^{-i \omega v}\mbox{e}^{i \omega r_{*}}(r-r_{0})^{\frac{i}{2\kappa_{0}}\omega}\nonumber\\
& = & \mbox{e}^{-i \omega v}(r-r_{0})^{\frac{i}{2\kappa_{0}}\omega}(r-r_{0})^{\frac{i}{2\kappa_{0}}\omega}\nonumber\\
& = & \mbox{e}^{-i \omega v}(r-r_{0})^{\frac{i}{\kappa_{0}}\omega}\ .
\label{eq:sol_out_2_canonical_Schw_tortoise}
\end{eqnarray}

Note that, if we put $a=Q=0$ into Eqs.~(96) and (97) of Ref. \cite{AnnPhys.350.14}, the resulting solutions are analyticly analogous to solutions given by Eqs.~(\ref{eq:sol_in_1_canonical_Schw_tortoise}) and (\ref{eq:sol_out_2_canonical_Schw_tortoise}).
%
%
\subsection{Analytic extension and radiation spectrum}
Now, we obtain by analytic continuation a real damped part of the outgoing wave solution of the massless scalar field which will be used to construct an explicit expression for the decay rate $\Gamma_{0}$. This real damped part corresponds (at least in part) to the temporal contribution to the decay rate \cite{IntJModPhysA.25.1463} found by the tunneling method used to investigate the analogue Hawking radiation.

From Eq.~(\ref{eq:sol_out_2_canonical_Schw_tortoise}), we see that this solution is not analytical in the acoustic event horizon, $r=r_{0}$. By analytic continuation, rotation by an angle $-\pi$ through the lower-half complex $r$ plane, give us
\begin{equation}
(r-r_{0}) \rightarrow \left|r-r_{0}\right|\mbox{e}^{-i\pi}=(r_{0}-r)\mbox{e}^{-i\pi}\ .
\label{eq:rel_3_canonical_Schw}
\end{equation}
Thus, the outgoing wave solution on the acoustic horizon surface $r_{0}$ is
\begin{equation}
\Psi_{out}(r<r_{0})=\mbox{e}^{-i\omega v}(r_{0}-r)^{\frac{i}{\kappa_{0}}\omega}\mbox{e}^{\frac{\pi}{\kappa_{0}}\omega}\ .
\label{eq:sol_1_out_4_canonical_Schw}
\end{equation}
Equations (\ref{eq:sol_out_2_canonical_Schw_tortoise}) and (\ref{eq:sol_1_out_4_canonical_Schw}) describe the outging wave outside and inside of the canonical acoustic black hole, respectively. Therefore, for an outgoing wave of a particle with energy $\omega > 0$, the outgoing decay rate or the relative scattering probability of the scalar wave at the acoustic event horizon surface, $r=r_{0}$, is given by
\begin{equation}
\Gamma_{0}=\left|\frac{\Psi_{out}(r>r_{0})}{\Psi_{out}(r<r_{0})}\right|^{2}=\mbox{e}^{-\frac{2\pi}{\kappa_{0}}\omega}\ ,
\label{eq:taxa_refl_canonical_Schw}
\end{equation}
which is a result already formally obtained in the literature \cite{IntJModPhysA.25.1463} in different context, and is analogous to the obtained in \cite{AnnPhys.350.14} for an astrophysical black hole.

According to the Damour-Ruffini-Sannan method \cite{PhysRevD.14.332,GenRelativGravit.20.239} for astrophysical black holes, a correct wave describing a particle flying off of the canonical acoustic black hole is given by
\begin{eqnarray}
\Psi_{\omega}(r) & = & N_{\omega}\ [\ H(r-r_{0})\ \Psi_{\omega}^{out}(r-r_{0})\nonumber\\
& + & H(r_{0}-r)\ \Psi_{\omega}^{out}(r_{0}-r)\ \mbox{e}^{\frac{\pi}{\kappa_{0}}\omega}\ ]\ ,
\label{eq:solucao_geral_onda_out_canonical_Schw}
\end{eqnarray}
where $\Psi_{\omega}^{out}(x)$ are the normalized wave functions given, from Eq.~(\ref{eq:sol_out_2_canonical_Schw_tortoise}), by
\begin{equation}
\Psi_{\omega}^{out}(x)=\mbox{e}^{-i \omega v}x^{\frac{i}{\kappa_{0}}\omega}\ .
\label{eq:sol_out_2_canonical_Schw_tortoise_x}
\end{equation}

Thus, from the normalization condition
\begin{equation}
\left\langle \Psi_{\omega}(r) | \Psi_{\omega}(r) \right\rangle=1=\left|N_{\omega}\right|^{2}\left[\mbox{e}^{\frac{2\pi}{\kappa_{0}}\omega}-1\right]\ ,
\label{eq:norm_onda_out_canonical_Schw}
\end{equation}
we get the resulting analogue Hawking radiation spectrum of scalar particles, which is given by
\begin{equation}
\left|N_{\omega}\right|^{2}=\frac{1}{\mbox{e}^{\frac{2\pi}{\kappa_{0}}\omega}-1}=\frac{1}{\mbox{e}^{\frac{\hbar\omega}{k_{B}T_{0}}}-1}\ .
\label{eq:espectro_rad_canonical_Schw_2}
\end{equation}
%
%
\section{Conclusions}
In this paper, we presented analytic solutions for radial part of the Klein-Gordon equation for a massless scalar field in the both rotating and canonical acoustic black holes. These general solutions are analytic solutions for all spacetime, which means, in the region between the acoustic event horizon and infinity. The radial solution is given in terms of the confluent Heun functions, and is valid over the range $0 \leq z < \infty$.

The obtained results for the rotating acoustic black hole have the advantage, as compared with the one obtained in literature \cite{ClassQuantumGrav.20.2929}, that the solutions are valid from the exterior event horizon to infinity, instead of to be valid only close to the exterior event horizon or at infinity. Otherwise, our results are valid for any frequency, and not for a restricted range of frequency, as presented in the literature \cite{PhysLettB.617.174}.

From these analytic solutions, we obtained the solutions for ingoing and outgoing waves near the acoustic horizon of a both rotating and canonical acoustic black holes, and used these results to discuss the Hawking radiation effect, in which we considered the properties of the confluent Heun functions to obtain the results. This approach has the advantage that it is not necessary the introduction of any coordinate system, as for example, the tortoise or Eddington-Finkelstein coordinates \cite{JHighEnergyPhys.07.070,IntJModPhysA.25.1463,PhysLettB.698.438}.

Generalizing the classical Damour-Ruffini method, we discussed the analogue Hawking radiation of both acoustic black holes. The expressions for the particle outgoing rates, given by Eqs.~(\ref{eq:taxa_refl_draining_bathtub_Kerr}) and (\ref{eq:taxa_refl_canonical_Schw}), describe the phenomena related to the radiation process for the rotating and canonical acoustic black holes, respectively.

As a final comment, we can say that we derived not only the Hawking temperature, but also, the Hawking black body spectrum for both acoustic black holes. This means that the rotating and canonical acoustic black holes behave not merely as thermal bodies but as black bodies.
%
%
\begin{acknowledgements}
The authors would like to thank Conselho Nacional de Desenvolvimento Cient\'{i}fico e Tecnol\'{o}gico (CNPq) for partial financial support.
\end{acknowledgements}
%
%

%
%
\section{Correction to: Acoustic black holes: massless scalar field analytic solutions and analogue Hawking radiation}
%
%
In Sect. 2, there is a mistake when we transformed Eq.~(16) into Eq.~(21), by using the new coordinate given by Eq.~(17) and its consequences expressed in Eqs.~(18)-(20). Correcting this mistake, the equation corresponding to Eq.~(21) should read
$$
\frac{d^{2}R}{dx^{2}}+\biggl(\frac{1}{x-x_{h}}-\frac{1}{2x}\biggr)\frac{dR}{dx}
$$
$$
+\biggl[\frac{m^{2}(B^{2}+c^{2}x_{h})}{4 c^{2} x_{h}^{2}}\frac{1}{x}+\frac{-B^{2}m^{2}-c^{2}m^{2}x_{h}+c^{2}x_{h}^{2}\omega^{2}}{4 c^{2} x_{h}^{2}}\frac{1}{x-x_{h}}
$$
$$
+\frac{(c x_{h} \omega-B m)^{2}}{4 c^{2} x_{h}}\frac{1}{(x-x_{h})^{2}}\biggr]R=0,
$$
where we used another relation for the new radial coordinate instead of Eq.~(17), given by
$$
x=r^{2}
$$
in such a way to obtain an equation which is identified with a confluent Heun equation. Now, using this new coordinate, Eq.~(18) is written as
$$
\Delta=x-\frac{A^{2}}{c^{2}}.
$$
Thus, the new horizon surface equation of the rotating acoustic black hole given by Eq.~(19) should read
$$
\Delta=x-x_{h}=0,
$$
and Eq.~(20) turns into
$$
x_{h}=r_{h}^{2}=\frac{A^{2}}{c^{2}}
$$
which corresponds to the new event horizon of the rotating acoustic black hole. In what follows, let us do homographic substitution given by Eq.~(22). Thus, Eq.~(23) should read
$$
\frac{d^{2}R}{dz^{2}}+\biggl(\frac{1}{z}+\frac{-1/2}{z-1}\biggr)\frac{dR}{dz}
$$
$$
+\biggl\{\frac{B^{2}m^{2}+c^{2}m^{2}x_{h}-c^{2}x_{h}^{2}\omega^{2}}{4 c^{2} x_{h}}\frac{1}{z}+\frac{-m^{2}(B^{2}+c^{2}x_{h})}{4 c^{2} x_{h}}\frac{1}{z-1}
$$
$$
-\biggl[i\biggl(\frac{c x_{h} \omega-B m}{2 c \sqrt{x_{h}}}\biggr)\biggr]^{2}\frac{1}{z^{2}}\biggr\}R=0.
$$
The \textit{F-homotopic transformation} of the dependent variable, $R(z) \mapsto U(z)$, is such that
$$
R(z)=z^{A_{1}}U(z),
$$
where the coefficient $A_{1}$ given by Eq.~(25) is now expressed as
$$
A_{1}=i\biggl(\frac{c x_{h} \omega-B m}{2 c \sqrt{x_{h}}}\biggr).
$$
In this case, Eq.~(26) should be changed to
$$
\frac{d^{2}U}{dz^{2}}+\biggl(\frac{2A_{1}+1}{z}+\frac{-1/2}{z-1}\biggr)\frac{dR}{dz}
$$
$$
+\biggl[\frac{B^2 m^2-i B c m \sqrt{h_{h}}+c^2 m^2 x_{h}+i c^{2} x_{h}^{3/2} \omega-c^2 x_{h}^2 \omega ^2}{4 c^2 x_{h}}\frac{1}{z}
$$
$$
-\frac{B^2 m^2-i B c m \sqrt{h_{h}}+c^2 x_{h}(m^{2}+i\sqrt{x_{h}} \omega)}{4 c^2 x_{h}}\frac{1}{z-1}\biggr]R=0.
$$
This functional form of the general solution of the radial part of the Klein-Gordon equation for a massless scalar field in the spacetime of a rotating acoustic black hole, in the exterior region of the acoustic event horizon, is similar to Eq.~(30), and therefore, its solution can be written in terms of the confluent Heun function as
$$
R(z)=z^{\frac{\beta}{2}}\{C_{1}\ \mbox{HeunC}(\alpha,\beta,\gamma,\delta,\eta;z)
$$
$$
+C_{2}\ z^{-\beta}\ \mbox{HeunC}(\alpha,-\beta,\gamma,\delta,\eta;z)\},
$$
where $C_{1}$ and $C_{2}$ are constants, and the parameters $\alpha$, $\beta$, $\gamma$, $\delta$, and $\eta$, which given originally by Eqs.~(31)-(35), are now given by the following expressions
$$
\alpha=0;
$$
$$
\beta=\frac{i(c x_{h} \omega - B m)}{c \sqrt{x_{h}}};
$$
$$
\gamma=-\frac{3}{2};
$$
$$
\delta=-\frac{x_{h} \omega ^2}{4};
$$
$$
\eta=\frac{1}{4} \biggl[3-m^{2}\biggl(1+\frac{B^{2}}{c^{2}x_{h}}\biggr)+x_{h}\omega^{2}\biggr].
$$
As a conclusion, the solution of the radial part of the Klein-Gordon equation is formally the same as Eq.~(30). In fact $\alpha$ assumes the same value and $\delta$ is given, formally, by the same expression as shown in Eq.~(34). As a consequence of the fact that the solution is the same, the analogue Hawking radiation, given by Eqs.~(37)-(60) are all correct and therefore, should be preserved.
%
%

In Sect. 3, a similar mistake was done when we transformed Eq.~(71) into Eq.~(78), using the coordinate transformation given by Eq.~(72). Now, let us correct this mistake. In order to do this, let write the line element which describes a canonical acoustic black hole in a more appropriate form given by
$$
ds^{2}=-c^{2}\Delta\ d\tau^{2}+\Delta^{-1}\ dr^{2}+r^{2}(d\theta^{2}+\sin^{2}\theta\ d\phi^{2}),
$$
where in this case Eq.~(65) should read
$$
\Delta=1-\frac{r_{0}^{4}}{r^{4}}.
$$
The covariant Klein-Gordon equation in the spacetime of a canonical acoustic black hole, given by Eq.~(69), is now written as
$$
\biggl[-\frac{r^{2}}{c^{2}\Delta}\frac{\partial^{2}}{\partial \tau^{2}}+\frac{\partial}{\partial r}\biggl(r^{2}\Delta\frac{\partial}{\partial r}\biggr)+\frac{1}{\sin\theta}\frac{\partial}{\partial\theta}\biggl(\sin\theta\frac{\partial}{\partial\theta}\biggr)+\frac{1}{\sin^{2}\theta}\frac{\partial^{2}}{\partial\phi^{2}}\biggr]\Psi=0.
$$
Thus, substituting Eq.~(70) into Eq.~(69), we find that Eq.~(71) should be written as
$$
\frac{d}{dr}\biggl(r^{2}\Delta\frac{dR}{dr}\biggr)+\biggl(\frac{\omega^{2}r^{2}}{c^{2}\Delta}-\lambda_{lm}\biggr)R=0.
$$
In order to obtain an equation which should be identified with one of the types of the Heun equation let us introduced a new coordinate transformation corresponding to Eq.~(72), given as follows
$$
x=r^{2}.
$$
Thus, using this new coordinate, Eq.~(73) turns into
$$
\Delta=1-\frac{r_{0}^{4}}{x^{2}}.
$$
Therefore, Eq.~(75) is simply written as
$$
\Delta=(x-x_{+})(x-x_{-})=0,
$$
whose solutions are the same given by Eqs.~(76)-(77), namely,
$$
x_{+}=+x_{0}=+r_{0}^{2},
$$
$$
x_{-}=-x_{0}=-r_{0}^{2},
$$
and correspond to the new acoustic event horizons of the canonical acoustic black hole. Thus, Eq.~(78) should read
$$
\frac{d^{2}R}{dx^{2}}+\biggl(\frac{1}{x-x_{+}}+\frac{1}{x-x_{-}}+\frac{3/2}{x}\biggr)\frac{dR}{dx}
$$
$$
+\biggl[\frac{A_{1}}{x-x_{+}}+\frac{A_{2}}{x-x_{-}}+\frac{A_{3}}{x}+\frac{B_{1}^{2}}{(x-x_{+})^{2}}+\frac{B_{2}^{2}}{(x-x_{-})^{2}}+\frac{B_{3}^{2}}{x^{3}}\biggr]R=0.
$$
From this point on, it is worth calling attention to the fact that the functional form of the radial part of the Klein-Gordon equation, for a massless scalar field in the spacetime of a canonical acoustic black hole, has changed. Then, we need to define some new coefficients, $A_{1}$, $A_{2}$, $A_{3}$, $B_{1}$, $B_{2}$, and $B_{3}$ which are given by
$$
A_{1}=\frac{-c^{2} \lambda ( x_{+}^{2} + x_{-}^{2})+2 c^{2} \lambda  x_{+} x_{-}+(x_{-}-3 x_{+}) \omega ^{2}}{4 c^{2} x_{+}^{2} (x_{+}-x_{-})^3},
$$
$$
A_{2}=\frac{c^{2} \lambda  (x_{+}^{2}+x_{-}^{2})-2 c^{2} \lambda  x_{+} x_{-}-(x_{+}-3 x_{-}) \omega ^{2}}{4 c^{2} x_{-}^{2} (x_{+}-x_{-})^3},
$$
$$
A_{3}=\frac{-c^{2} \lambda ( x_{+} + x_{-})+\omega ^{2}}{4 c^{2} x_{+}^{2} x_{-}^{2}},
$$
$$
B_{1}=\frac{\omega}{2 c \sqrt{x_{+}} (x_{+}-x_{-})},
$$
$$
B_{2}=\frac{\omega}{2 c \sqrt{x_{-}} (x_{+}-x_{-})},
$$
$$
B_{3}=\frac{i}{2}\sqrt{\frac{\lambda }{x_{+} x_{-}}}.
$$
Now, by setting the homographic substitution given by Eq.~(79), Eq.~(80) is now written as
$$
\frac{d^{2}R}{dz^{2}}+\biggl(\frac{1}{z}+\frac{1}{z-1}+\frac{3/2}{z-a}\biggl)\frac{dR}{dz}
$$
$$
+\biggl[\frac{-A_{1}x_{+}/a}{z}+\frac{-A_{2}x_{+}/a}{z-1}+\frac{-A_{3}x_{+}/a}{z-a}
$$
$$
+\frac{B_{1}^{2}}{z^{2}}+\frac{B_{2}^{2}}{(z-1)^{2}}+\frac{B_{3}^{2}}{(z-a)^{2}}\biggr]R=0,
$$
where the singular point $a$ is given by
$$
a=\frac{0-a_{1}}{a_{2}-a_{1}}=\frac{-x_{+}}{x_{-}-x_{+}}.
$$
Thus, the \textit{F-homotopic transformation} of the dependent variable, $R(z) \mapsto U(z)$, given by Eq.~(81), should read, in this case, as
$$
R(z)=z^{-\frac{1}{2}}(z-1)^{-\frac{1}{2}}(z-a)^{-\frac{3}{4}}U(z).
$$
Therefore, Eq.~(84) is rewritten as
$$
\frac{d^{2}U}{dz^{2}}+\biggl\{\frac{(1+4B_{1}^{2})/4}{z^{2}}+\frac{(1+4B_{2}^{2})/4}{(z-1)^{2}}+\frac{(3+16B_{3}^{2})/16}{(z-a)^{2}}
$$
$$
+\frac{(A_{1}+A_{3})x_{+}+a[-2+(A_{1}+A_{2})x_{+}]}{a}\frac{1}{(z-1)(z-a)}
$$
$$
+\frac{3+2a-4A_{1}x_{+}}{4}\frac{1}{z(z-1)(z-a)}\biggr\}R=0.
$$
Now, let us consider the general Heun equation, whose canonical form is [59]
$$
\frac{d^{2}y}{dz^{2}}+\biggl(\frac{\gamma}{z}+\frac{\delta}{z-1}+\frac{\epsilon}{z-a}\biggr)\frac{dy}{dz}+\frac{\alpha\beta z-q}{z(z-1)(z-a)}y=0,
$$
where $y(z)=\mbox{HeunG}(a,q;\alpha,\beta,\gamma,\delta;z)$ is the general Heun function. This is a Fuchsian type equation with regular singularities at $z=(0,1,a,\infty)$. The parameters $\alpha$, $\beta$, $\gamma$, $\delta$, $\epsilon$, $q$, $a$ are generally complex, arbitrary (except that $a \neq 0,1$), and related by
$$
\gamma+\delta+\epsilon=\alpha+\beta+1.
$$
Its normal form is given by
$$
\frac{d^{2}U}{dz^{2}}+\biggl\{\frac{(2\gamma-\gamma^{2})/4}{z^{2}}+\frac{(2\delta-\delta^{2})/4}{(z-1)^{2}}+\frac{(2\epsilon-\epsilon^{2})/4}{(z-a)^{2}}
$$
$$
+\frac{2\alpha\beta-\delta\epsilon-\gamma(\delta+\epsilon)}{2}\frac{1}{(z-1)(z-a)}
$$
$$
+\frac{\gamma(a\delta+\epsilon)-2q}{2}\frac{1}{z(z-1)(z-a)}\biggr\}R=0,
$$
where
$$
y(z)=z^{-\frac{\gamma}{2}}(z-1)^{-\frac{\delta}{2}}(z-a)^{-\frac{\epsilon}{2}}U(z).
$$
Thus, the general solution of the radial part of the Klein-Gordon equation for a massless scalar particle in the spacetime of a canonical acoustic black hole, in the exterior region of the acoustic event horizon, has changed and hence Eq.~(87) should read
$$
R(z) = z^{\frac{1}{2}(\gamma-1)}(z-1)^{\frac{1}{2}(\delta-1)}(z-a)^{\frac{1}{2}(\epsilon-2)}\{C_{1}\ \mbox{HeunG}(a,q;\alpha,\beta,\gamma,\delta;z)
$$
$$
+C_{2}\ z^{1-\gamma}\ \mbox{HeunG}(a,q_{1};\alpha_{1},\beta_{1},\gamma_{1},\delta;z)\},
$$
where $C_{1}$ and $C_{2}$ are constants, and the parameters $\alpha$, $\beta$, $\gamma$, $\delta$, $\epsilon$, and $q$, which are given by Eqs.~(88)-(92), are now given as follows
$$
\alpha = \frac{1}{4a}\{\sqrt{25 a^2-16 a^2( A_{1} x_{+}+ A_{2} x_{+}+ B_{1}^2+ B_{2}^2+B_{3}^2)-16 a x_{+} (A_{1} + A_{3})}
$$
$$
+a [4 i (B_{1}+ B_{2})+ 4+\sqrt{1-16 B_{3}^2}]\};
$$
$$
\beta=\frac{1}{2} [4 i (B_{1}+ B_{2})+\sqrt{1-16 B_{3}^2}+4]-\alpha;
$$
$$
\gamma=1+2iB_{1};
$$
$$
\delta=1+2iB_{2};
$$
$$
\epsilon=1+\frac{1}{2}\sqrt{1-16B_{3}^{2}};
$$
$$
q = \frac{1}{4} \{2 i B_{1} [a (2+4 i B_{2})+\sqrt{1-16 B_{3}^2}+2]+4 i a B_{2}+4 A_{1} x_{+}
$$
$$
+\sqrt{1-16 B_{3}^2}-1\}.
$$
The parameters $\alpha_{1}$, $\beta_{1}$, $\gamma_{1}$, and $q_{1}$ are given by the relations
$$
\alpha_{1}=\alpha+1-\gamma,
$$
$$
\beta_{1}=\beta+1-\gamma,
$$
$$
\gamma_{1}=2-\gamma.
$$
$$
q_{1}=q+(\alpha\delta+\epsilon)(1-\gamma).
$$
Thus, the solution of the radial part of the Klein-Gordon equation has changed. It is no more given by a confluent Heun function, but now it is given by a general Heun function. As a consequence, the analogue Hawking radiation is modified. Then, in order to explicit this modification, we need to discuss the expansion of the general Heun function, as follows. If $\gamma \neq \{0,-1,-2,...\}$, then from the Fuchs-Frobenius Theory, it follows that $\mbox{HeunG}(a,q;\alpha,\beta,\gamma,\delta;z)$ exists, is analytic in the disk $|z| < 1$, corresponds to exponent $0$ at $z=0$ and assumes the value $1$ there, and has the Maclaurin expansion
$$
\mbox{HeunG}(a,q;\alpha,\beta,\gamma,\delta;z)=\sum_{j=0}^{\infty}b_{j}z^{j},
$$
where $b_{0}=1$, and
$$
	a\gamma b_{1}-qb_{0}=0,
$$
$$
	X_{j}b_{j+1}-(Q_{j}+q)b_{j}+P_{j}b_{j-1}=0, \quad j \geq 1,
$$
with
$$
	P_{j}=(j-1+\alpha)(j-1+\beta),
$$
$$
	Q_{j}=j[(j-1+\gamma)(1+a)+a\delta+\epsilon],
$$
$$
	X_{j}=a(j+1)(j+\gamma).
$$
Thus, when $r \rightarrow r_{+} \Rightarrow x \rightarrow x_{+} \Rightarrow z \rightarrow 0$, Eq.~(93) should have been
$$
R(r) \sim C_{1}\ (r-r_{+})^{\frac{1}{2}(\gamma-1)}+C_{2}\ (r-r_{+})^{-\frac{1}{2}(\gamma-1)}\ .
$$
Then, considering the time factor, near the canonical acoustic black hole event horizon $r_{0}$, Eq.~(94) should be written as
$$
\Psi=\mbox{e}^{-i \omega t}(r-r_{-})^{\pm\frac{1}{2}(\gamma-1)}\ ,
$$
and thus Eq.~(95) should have be done by
$$
\frac{1}{2}(\gamma-1)=\frac{i}{2\kappa_{0}}\omega\ ,
$$
where
$$
\kappa_{0}=2cr_{0}^{3}.
$$
From this point on, the analysis of the analogue Hawking radiation expressed by Eqs.~(97)-(111) are all correct.

We apologize for the mistakes and express our gratitude to the colleagues for the comprehension.
%
%
\begin{acknowledgements}
The authors would like to thank Conselho Nacional de Desenvolvimento Cient\'{i}fico e Tecnol\'{o}gico (CNPq) for partial financial support. H. S. V. is funded through the research Project No. 150640/2018-8. V. B. B. is partially supported through the research Project No. 305835/2016-5. The authors also would like to thank Prof. Lu\'{i}s C. B. Crispino for the fruitful discussions.
\end{acknowledgements}

\begin{thebibliography}{99}
\bibitem{PhysRevLett.46.1351} W. G. Unruh, Phys. Rev. Lett. \textbf{46}, 1351 (1981).
\bibitem{CommunMathPhys.43.199} S. W. Hawking, Commun. Math. Phys. \textbf{43}, 199 (1975).
\bibitem{PhysRevD.82.044013} R. Banerjee, C. Kiefer and B. R. Majhi, Phys. Rev. D \textbf{82}, 044013 (2010)
\bibitem{PhysLettB.692.61} K. Umetsu, Phys. Lett. B \textbf{692}, 61 (2010).
\bibitem{PhysLettB.697.398} A. Yale, Phys. Lett. B \textbf{697}, 398 (2011).
\bibitem{EurophysLett.109.60006} H. S. Vieira, V. B. Bezerra and A. A. Costa, Europhys. Lett. \textbf{109}, 60006 (2015).
\bibitem{LivingRevRelativity.8.12} C. Barcel\'{o}, S. Liberati and M. Visser, Living Rev. Relativity \textbf{8}, 12 (2005).
\bibitem{NewJPhys.10.053015} G. Rousseaux, C. Mathis, P. Ma\"{i}ssa, T. G. Philbin and U. Leonhardt, New J. Phys. \textbf{10}, 053015 (2008).
\bibitem{PhysRevLett.106.021302} S. Weinfurtner, E. W. Tedford, M. C. J. Penrice, W. G. Unruh and G. A. Lawrence, Phys. Rev. Lett. \textbf{106}, 021302 (2011).
\bibitem{PhysRevD.90.044033} F. Michel and R. Parentani, Phys. Rev. D \textbf{90}, 044033 (2014).
\bibitem{PhysRevLett.85.4643} L. J. Garay, J. R. Anglin, J. I. Cirac and P. Zoller, Phys. Rev. Lett. \textbf{85}, 4643 (2000).
\bibitem{PhysRevLett.91.240407} P. O. Fedichev and U. R. Fischer, Phys. Rev. Lett. \textbf{91}, 240407 (2003).
\bibitem{NewJPhys.13.063048} I. Zapata, M. Albert, R. Parentani and F. Sols, New J. Phys. \textbf{13}, 063048 (2011).
\bibitem{PhysRevLett.105.240401} O. Lahav, A. Itah, A. Blumkin, C. Gordon, S. Rinott, A. Zayats and Jeff Steinhauer, Phys. Rev. Lett. \textbf{105}, 240401 (2010).
\bibitem{PhysRevD.58.064021} T. A. Jacobson and G. E. Volovik, Phys. Rev. D \textbf{58}, 064021 (1998).
\bibitem{JHighEnergyPhys.06.087} X. H. Ge and S. J. Sin, J. High Energy Phys. \textbf{06}, 087 (2010).
\bibitem{IntJModPhysD.21.1250038} X. H. Ge, S. F. Wu, Y. Wang, G. H. Yang and Y. G. Shen, Int. J. Mod. Phys. D \textbf{21}, 1250038 (2012).
\bibitem{PhysRevB.86.144505} D. Gerace and I. Carusotto, Phys. Rev. B \textbf{86}, 144505 (2012).
\bibitem{PhysRevLett.94.061302} S. Giovanazzi, Phys. Rev. Lett. \textbf{94}, 061302 (2005).
\bibitem{PhysRevLett.107.149401} R. Sch\"{u}tzhold and W. G. Unruh, Phys. Rev. Lett. \textbf{107}, 149401 (2011).
\bibitem{PhysRevLett.107.149402} F. Belgiorno, S. L. Cacciatori, M. Clerici, V. Gorini, G. Ortenzi, L. Rizzi, E. Rubino, V. G. Sala and D. Faccio, Phys. Rev. Lett. \textbf{107}, 149402 (2011).
\bibitem{PhysRevD.85.084014} S. Liberati, A. Prain and M. Visser, Phys. Rev. D \textbf{85}, 084014 (2012).
\bibitem{EurPhysJPlus.127.78} S. Finazzi and I. Carusotto, Eur. Phys. J. Plus \textbf{127}, 78 (2012).
\bibitem{PhysRevA.78.063804} F. Marino, Phys. Rev. A \textbf{78}, 063804 (2008).
\bibitem{PhysRevA.80.065802} F. Marino, M. Ciszak and A. Ortolan, Phys. Rev. A \textbf{80}, 065802 (2009).
\bibitem{EurophysLett.92.14002} I. Fouxon, O. V. Farberovich, S. Bar-Ad and V. Fleurov, Europhys. Lett. \textbf{92}, 14002 (2010).
\bibitem{PhysRevB.84.233405} D. D. Solnyshkov, H. Flayac and G. Malpuech, Phys. Rev. B \textbf{84}, 233405 (2011).
\bibitem{NewJPhys.13.085005} E. Rubino, F. Belgiorno, S. L. Cacciatori, M. Clerici, V. Gorini, G. Ortenzi, L. Rizzi, V. G. Sala, M. Kolesik and D. Faccio, New J. Phys. \textbf{13}, 085005 (2011).
\bibitem{NaturePhys.10.864} J. Steinhauer, Nature Phys. \textbf{10}, 864 (2014).
\bibitem{PhysRevD.83.124016} A. Fabbri and C. Mayoral, Phys. Rev. D \textbf{83}, 124016 (2011).
\bibitem{PhysRevD.83.124047} C. Mayoral, A. Fabbri and M. Rinaldi, Phys. Rev. D \textbf{83}, 124047 (2011).
\bibitem{NewJPhys.10.103001} I. Carusotto, S. Fagnocchi, A. Recati, R. Balbinot and A. Fabbri, New J. Phys. \textbf{10}, 103001 (2011).
\bibitem{PhysRevD.92.024043} J. Steinhauer, Phys. Rev. D \textbf{92}, 024043 (2015).
\bibitem{ClassQuantumGrav.18.1137} C. Barcel\'{o}, S. Liberati and M. Visser, Class. Quantum Grav. \textbf{18}, 1137 (2001).
\bibitem{GenRelGrav.34.1719} M. Visser, C. Barcel\'{o} and S. Liberati, Gen. Rel. Grav. \textbf{34}, 1719 (2002).
\bibitem{JHighEnergyPhys.04.030} S. R. Das, A. Ghosh, J. H. Oh and A. D. Shapere, J. High Energy Phys. \textbf{04}, 030 (2011).
\bibitem{JPhysBAtMolOptPhys.45.163001} S. J. Robertson, J. Phys. B: At. Mol. Opt. Phys. \textbf{45}, 163001 (2012).
\bibitem{PhysRevA.78.021601} S. W\"{u}ster, Phys. Rev. A \textbf{78}, 021601(R) (2008).
\bibitem{PhysRevD.90.104015} A. Belenchia, S. Liberati and A. Mohd, Phys. Rev. D \textbf{90}, 104015 (2014).
\bibitem{PhysRevD.81.124013} E. S. Oliveira, S. R. Dolan and L. C. B. Crispino, Phys. Rev. D \textbf{81}, 124013 (2010).
\bibitem{PhysRevD.79.064014} S. R. Dolan, E. S. Oliveira and L. C. B. Crispino, Phys. Rev. D \textbf{79}, 064014 (2009).
\bibitem{PhysRevD.70.124006} E. Berti, V. Cardoso and J. P. S. Lemos, Phys. Rev. D \textbf{70}, 124006 (2004).
\bibitem{PhysRevD.51.2827} W. G. Unruh, Phys. Rev. D \textbf{51}, 2827 (1995).
\bibitem{PhysRevD.44.1731} T. Jacobson, Phys. Rev. D \textbf{44}, 1731 (1991).
\bibitem{PhysRevD.48.728} T. Jacobson, Phys. Rev. D \textbf{48}, 728 (1993).
\bibitem{ClassQuantumGrav.20.3907} S. Basak and P. Majumdar, Class. Quantum Grav. \textbf{20}, 3907 (2003).
\bibitem{ClassQuantumGrav.20.2929} S. Basak and P. Majumdar, Class. Quantum Grav. \textbf{20}, 2929 (2003).
\bibitem{PhysLettB.617.174} S. Lepe and J. Saavedra, Phys. Lett. B \textbf{617}, 174 (2005).
\bibitem{Slavyanov:2000} S. Y. Slavyanov and W. Lay, \textit{Special functions, A unified theory based on singularities}, (Oxford University Press, New York, 2000).
\bibitem{PhysRevD.13.2188} J. B. Hartle and S. W. Hawking, Phys. Rev. D \textbf{13}, 2188 (1976).
\bibitem{PhysRevD.15.2088} S. M. Christensen and S. A. Fulling, Phys. Rev. D \textbf{15}, 2088 (1977).
\bibitem{PhysRevLett.95.011303} S. P. Robinson and F. Wilczek, Phys. Rev. Lett. \textbf{95}, 011303 (2005).
\bibitem{PhysRevLett.85.5042} M. K. Parikh and F. Wilczek, Phys. Rev. Lett. \textbf{85}, 5042 (2000).
\bibitem{JHighEnergyPhys.07.070} W. Kim and H. Shin, J. High Energy Phys. \textbf{07}, 070 (2007).
\bibitem{IntJModPhysA.25.1463} R. B\'{e}car, P. Gonz\'{a}lez, G. Pulgar and J. Saavedra, Int. J. Mod. Phys. A \textbf{25}, 1463 (2010).
\bibitem{PhysLettB.698.438} L. C. Zhang, H. F. Li and R. Zhao, Phys. Lett. B \textbf{698}, 438 (2011).
\bibitem{PhysLettB.694.149} M. A. Anacleto, F. A. Brito and E. Passos, Phys. Lett. B \textbf{694}, 149 (2010).
\bibitem{ClassQuantumGrav.15.1767} M. Visser, Class. Quantum Grav. \textbf{15}, 1767 (1998).
\bibitem{Ronveaux:1995} A. Ronveaux, \textit{Heun's differential equations}, (Oxford University Press, New York, 1995).
\bibitem{JPhysAMathTheor.43.035203} P. P. Fiziev, J. Phys. A: Math. Theor. \textbf{43}, 035203 (2010).
\bibitem{AnnPhys.350.14} H. S. Vieira, V. B. Bezerra and C. R. Muniz, Ann. Phys. (NY) \textbf{350}, 14 (2014).
\bibitem{PhysRevD.14.332} T. Damour and R. Ruffini, Phys. Rev. D \textbf{14}, 332 (1976).
\bibitem{GenRelativGravit.20.239} S. Sannan, Gen. Relativ. Gravit. \textbf{20}, 239 (1988).
\bibitem{arXiv:9901047} M. Visser, arXiv:9901047 \textbf{[gr-qc]} (1999).
\bibitem{Arfken:2005} G. B. Arfken and H. J. Weber, \textit{Mathematical methods for physicists}, (Elsevier Academic Press, San Diego, 2005).
\end{thebibliography}
\end{document}